\documentclass{article}
\usepackage{newcent}
\usepackage{bifchaos}
\usepackage{multicol}
\usepackage{vmargin}
\usepackage[dvips]{graphicx}

\setpapersize{A4}
\setmarginsrb{1.3cm}{2cm}{2cm}{2.5cm}{0pt}{0mm}{0pt}{0mm}

\newfont{\Bbb}{msbm10}
\newcommand{\T}{\mbox{\Bbb T}}

\begin{document}

\title{\bf THREE-FREQUENCY RESONANCES \\ IN DYNAMICAL SYSTEMS}

\author{
OSCAR CALVO,\thanks{Email calvo@athos.fisica.unlp.edu.ar} 
\\
{\it CICpBA, 
L.E.I.C.I., Departamento de Electrotecnia, Facultad de Ingenier\'{\i}a,} \\   
{\it Universidad Nacional de La Plata, 1900 La Plata, Argentina} \\
\\
JULYAN H. E. CARTWRIGHT,\thanks{Email julyan@hp1.uib.es, WWW
http://formentor.uib.es/$\sim$julyan} 
\\
{\it Instituto Andaluz de Ciencias de la Tierra, IACT (CSIC-UGR),} \\
{\it E-18071 Granada, Spain} \\
\\
DIEGO L. GONZ\'ALEZ,\thanks{Email diego@indra.lamel.bo.cnr.it} 
\\
{\it Istituto Lamel, CNR, I-40129 Bologna, Italy} \\
\\
ORESTE PIRO,\thanks{Email piro@imedea.uib.es, WWW
http://www.imedea.uib.es/$\sim$piro} 
\\
{\it Institut Mediterrani d'Estudis Avan\c{c}ats, IMEDEA (CSIC--UIB),} \\
{\it E-07071 Palma de Mallorca, Spain} \\
\\
OSVALDO A. ROSSO,\thanks{Email rosso@ulises.ic.fcen.uba.ar}  
\\
{\it Instituto de C\'alculo, Facultad de Ciencias Exactas y Naturales,}  \\
{\it Universidad de Buenos Aires, 1428 Buenos Aires, Argentina}  \\
}

\date{Int. J. Bifurcation and Chaos, {\bf 9}, 2181--2187 (1999)}

\maketitle

\begin{abstract}
We investigate numerically and experimentally dynamical systems having three
interacting frequencies: a discrete mapping (a circle map), an exactly solvable
model (a system of coupled ordinary differential equations), and an
experimental device (an electronic oscillator). We compare the hierarchies of
three-frequency resonances we find in each of these systems. All three show
similar qualitative behaviour, suggesting the existence of generic features in
the parameter-space organization of three-frequency resonances.
\end{abstract}

\begin{multicols}{2}

\begin{figure*}[t]
\begin{center}
\includegraphics[width=\textwidth]{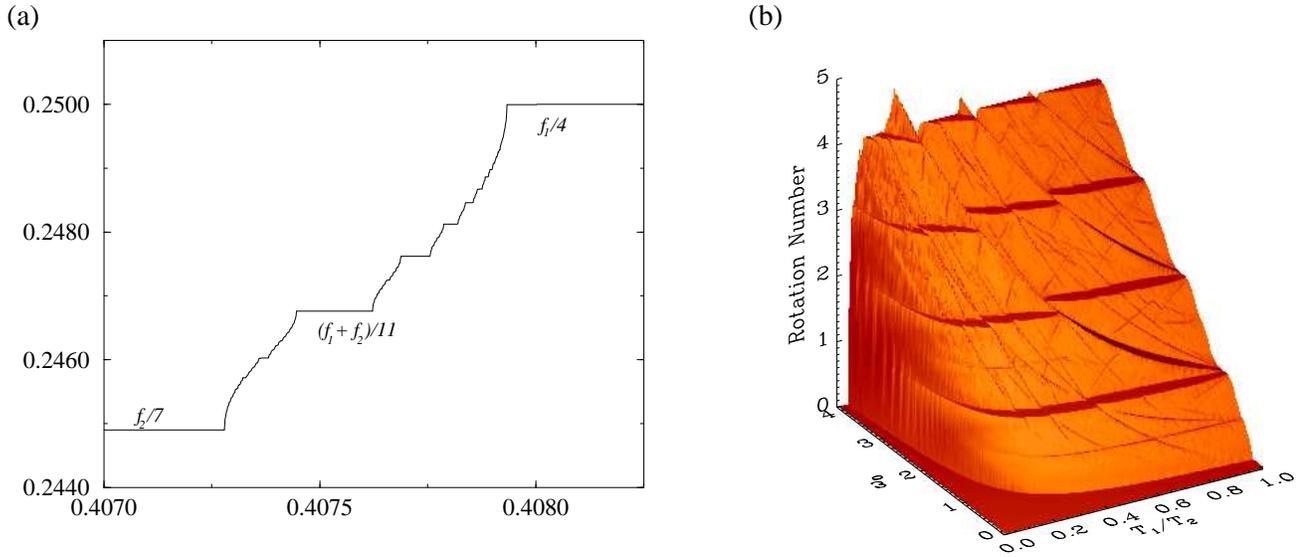}
\end{center}
\caption{
(a) Three-frequency devil's staircase in the quasiperiodically forced circle map.
The external frequencies are $f_1=1$ and $f_2=12/7$. 
The steps at left and right
correspond to the plateaux $(12/7)/7=0.2448979\ldots$ and $1/4=0.25$;
the largest plateau between the two is 
$(1+12/7)/11=(19/7)/11=0.2467532\ldots$, in accordance with the prediction of
the generalized Farey mediant.
(b) Devil's ramp system: the rotation number as a function of the external
frequency ratio and the intrinsic frequency for a quasiperiodically
forced circle map representing a doubly kicked nonlinear oscillator.
The devil's ramps illustrate the overall ordering of three-frequency resonances.
}\label{circle}
\end{figure*}

\section{Introduction} 

A fundamental goal in the modelling of complex behaviour by means of
low-dimensional nonlinear deterministic systems is the understanding of the
transition to chaos from quasiperiodic motion on a $\T^j$ torus for small $j$.
This transition is well understood for the case $j=2$, the paradigmatic model
being the periodically forced nonlinear oscillator and its discrete version: 
the circle map.

In $\T^2$, orbits can be hierarchically arranged by means of  rotation numbers
\cite{arrowsmith} 
\begin{equation}
\rho(r,\theta)=\lim_{n\to\infty}{\frac{\theta_n-\theta_0}{2\pi n}}
,\end{equation} 
periodic orbits --- lockings or resonances --- being then
characterized by rational rotation numbers. If two rational rotation numbers
$(p/q,r/s)$ satisfy $|ps-rq|=1$ they are said to be unimodular or Farey
adjacents. Between two periodic solutions characterized by unimodular rotation
numbers there exists a periodic solution with a minimal period. This solution
is given by a rotation number termed the Farey mediant of the two previous
ones: $m/n = p/q \oplus r/s = (p+r)/(q+s)$. Moreover this periodic solution is
the most prominent --- the largest --- in the open interval $(p/q,r/s)$ between
the two parents. Continued application of the mediant gives the Farey tree
which underlies the organization of the locking intervals of rational rotation
numbers in the space of parameters of two-frequency systems; the so-called
devil's staircase \cite{oreste,aronson,oreste2,cvit,hao,bogpaper}. These
properties of rotation numbers are closely related to the continued fraction
expansion which gives the best rational approximants to an arbitrary real
number. 

In order to study $\T^3$ torus breakdown we have generalized this approach to
the case of the simultaneous approximation of a pair of real numbers
\cite{3freq}. Three-frequency systems also possess a structure of resonances,
but this time more complex, for in addition to the rational relations found in
two-frequency systems there is now a new type of locking: a three-frequency
resonance, given by the nontrivial solutions of the equation 
$nf_1+mf_2+qf_3=0$ with $n$, $m$, and $q$ nonzero integers. 
Three-frequency resonances form a web in the parameter space of the 
frequencies \cite{baesens}.

We have found that \cite{3freq}:
\begin{itemize}
\item We can define a subharmonic real interval $(f_1/p,f_2/q)$ 
inside which the hierarchical organization of three-frequency resonances is
well described by a generalized Farey sum.
\item In order to implement the generalized Farey sum we must modify the
adjacency condition: if $p/q$ is a convergent of the forcing frequency ratio we
define two fractions of real numbers $f_x /m$, $f_y/n$ as adjacents 
if $|f_x n-f_y m|=|f_1 q-f_2 p|$ where $f_1$ and
$f_2$ are the external frequencies.
\item We can now define the generalized Farey sum between fractions of real
numbers which satisfy this adjacency condition: 
$f_r/k =f_x/m\oplus f_y/n=(f_x+f_y)/(m+n)$.
\end{itemize}

Here we present numerical and experimental studies of three dynamical systems
having three interacting frequencies, in which we can observe three-frequency
resonances, and compare our observations with the above predictions.

\begin{figure*}[t]
\begin{center}
\includegraphics[width=\textwidth]{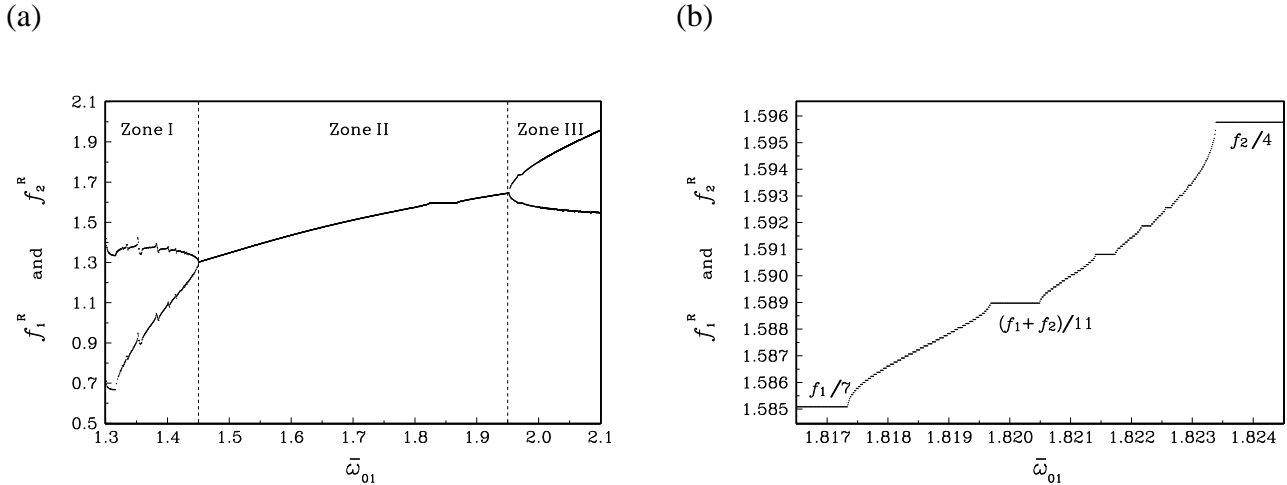}
\end{center}
\caption{(a) The frequency response of both oscillators of Eq.(\ref{pirogon})
as a function of the intrinsic frequency of the first oscillator. Zone I is 
chaotic, with intermittent synchronization, in zone II there is 1/1 
synchronization, and in zone III, synchronization other than 1/1. The input 
frequencies have been set to be equivalent to those of Fig.~\ref{circle}.
(b) A magnification of the central zone of (a), where the oscillators are 
mutually synchronized at 1/1. 
}\label{numosc}
\end{figure*}

\section{Numerical Simulations}

We have investigated a discrete mapping: a quasiperiodically forced circle 
map given by the equation
\begin{equation}
\phi'=\phi+\Omega_n-{\frac{k}{2\pi}}\sin 2\pi\phi\bmod 1
.\end{equation}
The quasiperiodic sequence $\Omega_n$ is the time interval between successive
pulses of a sequence composed of the superposition of two periodic
subsequences, one of period $T_1 = 1$ and the other of period $T_2$ ($T_2 < 1$
with no loss of generality), multiplied by the value of the intrinsic frequency
of the oscillator.

In Fig.~\ref{circle}a we can observe that there are intervals of constant
frequency response, namely the horizontal segments of the figure. These
intervals represent stable three-frequency resonances, that is, responses that
satisfy $nf_1+mf_2+qf_3=0$ with $n$, $m$, and $q$ nonzero integers. We can also
see in Fig.~\ref{circle}a that the stability widths of the resonances form a
hierarchical structure very similar to that of the well known devil's staircase
in periodically forced nonlinear oscillators. This is the generalized devil's
staircase for three-frequency systems.
 
The other image, Fig.~\ref{circle}b, shows the devil's ramps. These 
represent the
global hierarchy of three-frequency resonances shown as a function of the 
external frequency ratio and the intrinsic frequency for a critical ($k=1$, the
map instantaneously has critical behaviour) quasiperiodically forced circle map.

We have also performed numerical simulations with 
two parametrically coupled nonlinear oscillators, each with an exact analytical
solution, forced by means of two impulsive periodic forces. The
differential equation for each oscillator \cite{oreste,oreste2} is: 
\begin{equation}
\ddot u_i+(4bu_i^2-2a)\dot u_i+b^2u_i^5-2abu_i^3+(\omega_{0i}(t)^2+a^2)u_i
=f_{i}(t) 
\label{pirogon}\end{equation}
for $i=1,2$.
The coupling and forcing terms
$\omega_{0i}(t)=
\tilde\omega_{0i}+{\mathrm sgn}\,u_{i}(t)\,{\mathrm sgn}\,u_{j}(t)\Delta_i$   
and $f_{i}(t)=V_{Ei}\sum_n\delta(t-nT_{Ei})$
preserve the piecewise integrability of the system.

We have made a power spectrum analysis of the output of both oscillators. In
Fig.~\ref{numosc}a we display the most prominent peak in each spectrum versus
the intrinsic frequency of oscillator 1 for a parameter region  equivalent to
that of Fig.~\ref{circle}a. In Fig.~\ref{numosc}b we show a magnification of
the zone of Fig.~\ref{numosc}a in which the two oscillators are synchronized
at $1/1$. The three principal peaks in the Fourier spectrum satisfy
$nf_1+mf_2+qf_3=0$ with $n$, $m$, and $q$ nonzero integers. All the other peaks
in the spectrum can be expressed as linear combinations of this fundamental
set. 

\begin{figure*}[tb]
\begin{center}
\includegraphics[width=0.6\textwidth]{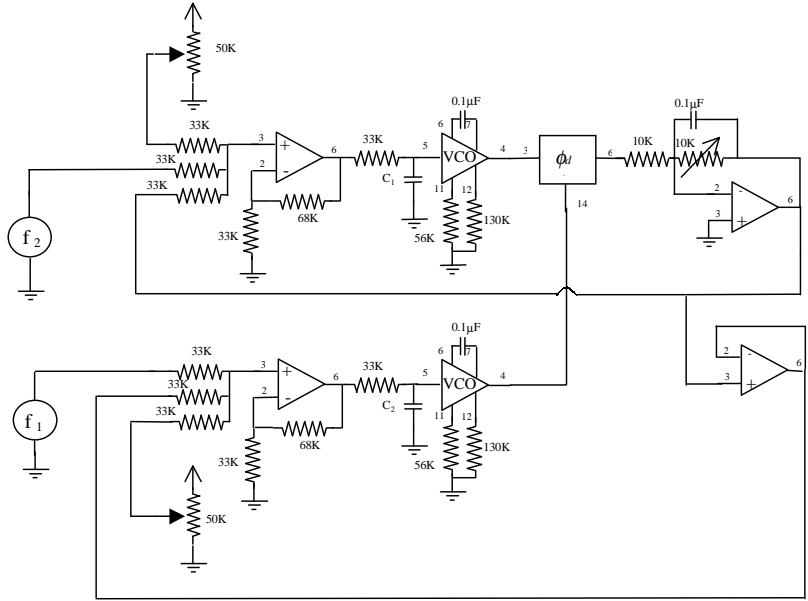}
\end{center}
\caption{The circuit.
The output of the voltage controlled oscillator (VCO) sections of both digital 
phase-locked loop devices are sent to an exclusive OR port for phase 
comparison. The phase comparator output is returned, after low-pass filtering,
to both VCO inputs. An amplifier in the feedback path controls the
coupling strength between the oscillators. Appropriate operational adders on 
both VCO inputs allow the external forcing of the oscillators. Stable 
phase-locked responses between oscillators also require an additional unit 
gain inverter prior to one of the device inputs.
}\label{circuit_diag}
\end{figure*}

\begin{figure*}[tb]
\begin{center}
\includegraphics[width=\textwidth]{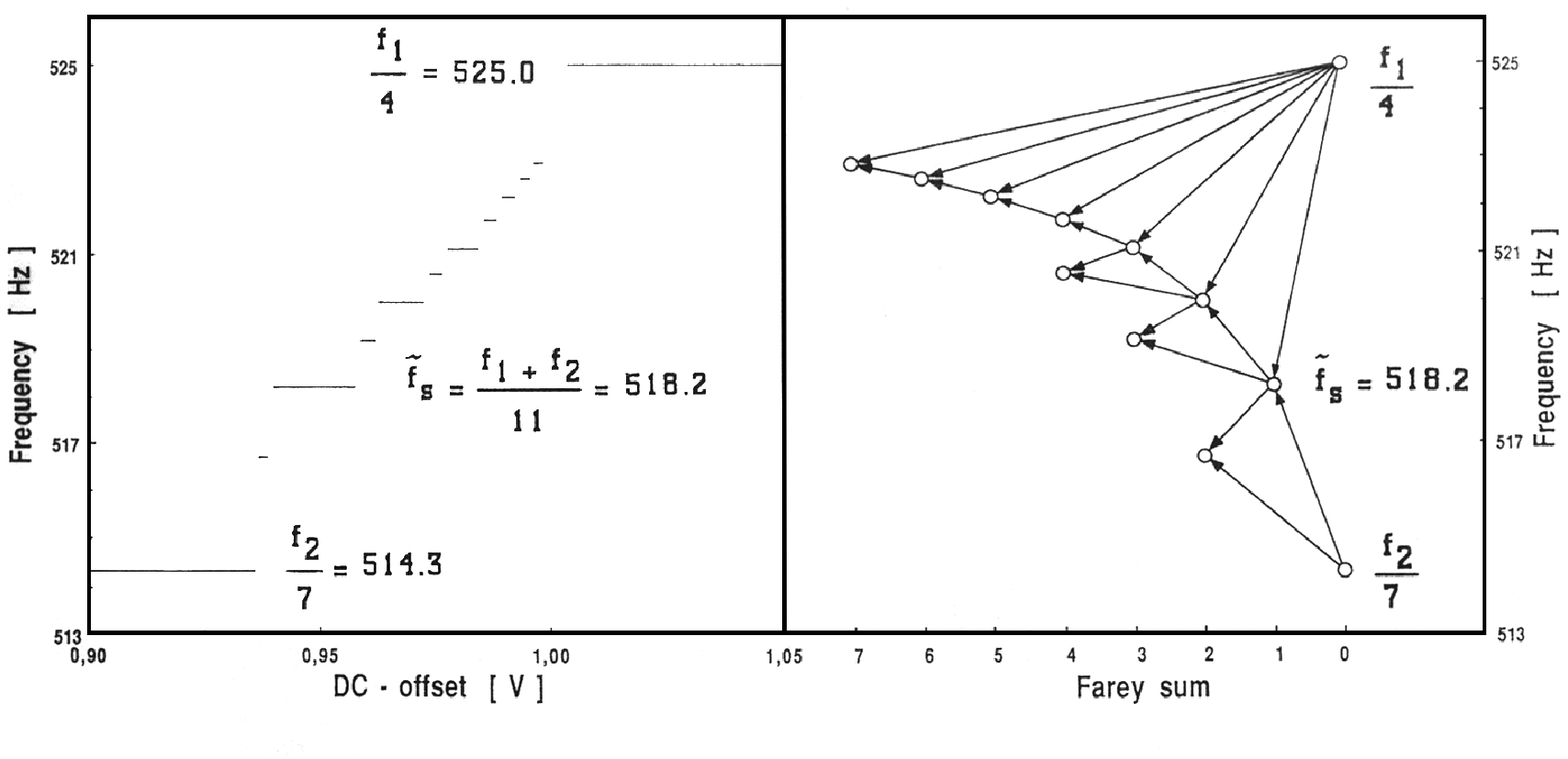}
\end{center}
\caption{
(a) Experimental results from an electronic circuit of quasiperiodically
forced phase-locked loops --- a three-frequency devil's staircase. The 
external frequencies $f_1$ and $f_2$ are here fixed at 2100 Hz and 3600 Hz,
equivalent to those in Fig.~\ref{circle}. We
have plotted the third frequency of a three-frequency resonance against
a control parameter (the DC offset of one of the external forces) for all
resonances with plateaux larger than a certain size. 
(b) Shows the
hierarchy of three-frequency resonances predicted by the generalized mediant
starting from the parents $f_1/4$ and $f_2/7$.  At each level in the
hierarchy, the daughter resonance formed by the mediant between two adjacent
parents is seen to be the largest in its interval.}\label{pll}
\end{figure*}

\section{Experimental Results}
We have constructed an electronic oscillator --- Fig.~\ref{circuit_diag} --- a
higher-dimensional version of a phase-locked loop (PLL), forced with two
independent periodic forces \cite{ourosc}. 
Our circuit consists of two coupled voltage-controlled oscillators forced with
two external forces of frequencies $f_1$ and $f_2$. As a basic
circuit for both oscillators we use a digital phase-locked loop integrated
circuit, the CD 4046A. The outputs of the two phase-locked loops are sent to a
type 1 phase comparator. The error signal is fed back to both
voltage-controlled oscillators, and passes through an overall adjustable
amplifier to provide control over the coupling strength; we are interested in
the weak coupling regime in this work. Inverted and direct versions of the
error signal are sent to oscillators one and two respectively; this inversion
of the error signal in one of the paths is necessary for the stability of the
circuit. Feedback signals enter the voltage-controlled-oscillator control pins
through appropriate adder circuits. The adders also allow independent coupling
with the external forces and tuning of the internal frequencies through
application of adjustable DC levels. 

\begin{figure*}[tb]
\begin{center}
\includegraphics[width=0.6\textwidth]{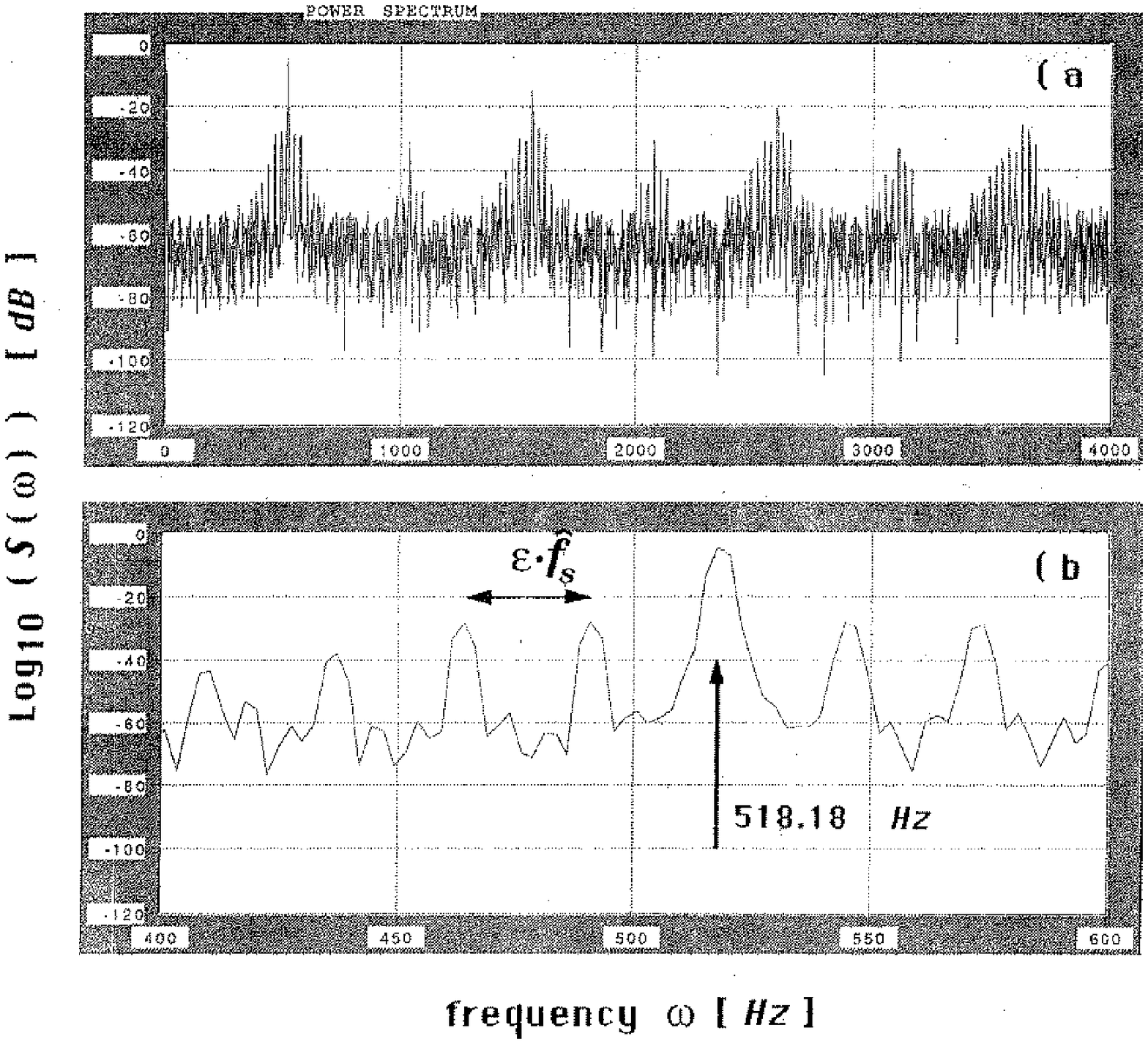}
\end{center}
\caption{The power spectrum of the output of the circuit for a DC offset of 0.95
V. Other parameters are as in fig.~\ref{pll}. (a) From 0 -- 4000 Hz.
(b) Detail from 400 -- 600 Hz, showing peak at $\tilde f_s\approx 518.18$ Hz
and minor peaks separated a distance $\delta=\varepsilon\tilde f_s$.}
\label{pwr_spect}
\end{figure*}

In Fig.~\ref{pll}a we plot the fundamental frequency $f_3$ versus the DC offset
of one of the forcing signals. As in the previous numerical simulations, the
generalized mediant is the largest plateau between the two adjacents given by
the subharmonics of the forcing frequencies. The circuit of
Fig.~\ref{circuit_diag} represents qualitatively a large class of real
dynamical systems characterized by some kind of feedback loop with a
transference function of the low-pass type, and can be useful in clarifying the
physical mechanisms that underlie three-frequency resonances. In 
Fig.~\ref{pwr_spect}a we show the Fourier spectrum of the oscillator response
at a point inside the stability region for the three-frequency resonance. In
Fig.~\ref{pwr_spect}b, a magnification around this frequency value, we can see
that harmonics of the main resonance at $\tilde f_s=518.2$ Hz lie at equal
distances $\delta=27.3$ Hz from both external frequencies. The peak at $\tilde
f_s$  dominates the spectrum, being greater than those corresponding to the
forcing frequencies. The distance $\delta=\varepsilon\tilde f_s$ appears as a
modulation of the spectrum, which shows, on this fine scale, a series of peaks
separated precisely by this distance. 

The origin of these frequencies can be interpreted in terms of the
general structure of the system. Passive nonlinearities are able to generate
appropriate frequency harmonics. These, in turn, through nonlinear mixing with
the driving frequencies, can generate the frequency $\delta$. 
Only for the case of the main three-frequency resonance are the two $\delta$s
of the same value, i.e., $\delta=\delta_1=\delta_2$. Thus, further nonlinear
mixing in this case gives only terms of zero frequency and harmonics of
$\delta$, that is, no other frequencies are added to the system. Otherwise, for
an arbitrary frequency, the two $\delta$s are in general different and, if
irrationally related, generate an infinity of low-frequency components that
pass through the low-pass filter and are fed back to the system, destabilizing
the response for this particular frequency  value. This physical mechanism can
also explain the ordering of the three-frequency resonances. Only the main
resonance is characterized by $\delta_1=\delta_2$, but successive application
of the generalized Farey operation gives distances that are rationally related,
that is  $\delta_1/\delta_2=r/s$. Consequently, for all resonances, successive
nonlinear mixing through the feedback loop can generate only a finite number of
new frequencies, preserving the stability of the corresponding response. 
Also the greater the integers $r$, $s$, the smaller the stability
interval, because more destabilizing frequencies are added through the feedback
mechanism.

\section{Discussion}

In Fig.~\ref{pll}b we plot the generalized Farey tree structure formed by the
predicted three-frequency resonances \cite{3freq}. The frequencies of the
resonances are obtained by recursive application of the generalized Farey sum
starting with $f_1/q$ and $f_2/p$. This structure accurately describes all the
three-frequency resonances found in the experiment with phase-locked loops
\cite{ourosc} and in the numerical simulations with differential equations and
maps. Successive levels in the tree describe the ordering of stability widths
in each case. The generalized Farey tree structure is thus found to govern the
hierarchy of three-frequency resonances in representative dynamical systems
with three interacting frequencies. We conjecture from this that the ordering
may be universal in three-frequency systems.

From the theorems of Newhouse, Ruelle, and Takens \cite{ruelle,newhouse} we
expect that this hierarchical structure of three-frequency resonances should be
relevant to the study of torus breakdown and the transition to chaos in complex
arrays of coupled nonlinear oscillators. Such oscillator networks occur in many
biological systems, from fireflies and circadian rhythms to physiological and
neurological systems such as the heart and brain. We have investigated the
application of these ideas to one such problem in biology: that of the
mechanism of pitch perception in the auditory system. We find good agreement
between dynamical systems theory and perceptual experiments
\cite{soundsletterprl}.

\section*{Acknowledgements}
JHEC and OP acknowledge the financial support of the Spanish Direcci\'on 
General de Investigaci\'on Cient\'\i fica y T\'ecnica, contracts PB94-1167 and 
PB94-1172.

\bibliographystyle{bifchaos} 
\bibliography{database}

\end{multicols}

\end{document}